\documentclass[12pt,preprint]{aastex}

\shorttitle{The Scintillating Source PKS~1257$-$326}

\begin{document}

\title{Rapid Variability and Annual Cycles in the Characteristic
Time-scale of the Scintillating Source PKS~1257$-$326}

\author{H. E. Bignall,\altaffilmark{1,2}
 D. L. Jauncey,\altaffilmark{2} 
 J. E. J. Lovell,\altaffilmark{2}
 A. K. Tzioumis,\altaffilmark{2}   
 L. Kedziora-Chudczer,\altaffilmark{2,3}
 J.-P. Macquart,\altaffilmark{4}
 S. J. Tingay,\altaffilmark{2}
 D. P. Rayner,\altaffilmark{2}
 and R. W. Clay\altaffilmark{1}}

\altaffiltext{1}{Department of Physics and Mathematical Physics,
University  of Adelaide, SA 5005, Australia;
hbignall@physics.adelaide.edu.au}

\altaffiltext{2}{CSIRO Australia Telescope National Facility, P.O. Box
76,  Epping, NSW 1710, Australia}

\altaffiltext{3}{Anglo-Australian Observatory, Epping, NSW, Australia}

\altaffiltext{4}{Kapteyn Institute, University of Groningen, The
Netherlands}

\begin{abstract}

Rapid radio intra-day variability (IDV) has been discovered in the
southern quasar PKS~1257$-$326. Flux density changes of up to 40\% in
as little as 45 minutes have been observed in this source,  making it,
along with PKS~0405$-$385 and J1819$+$3845, one of the three most
rapid IDV sources known. We have monitored the IDV in this source with
the Australia Telescope Compact Array (ATCA) at 4.8 and 8.6\,GHz over
the course of the last year, and find a clear {\it annual cycle} in
the characteristic time-scale of variability. This annual cycle
demonstrates unequivocally that interstellar scintillation is the
cause of the rapid IDV at radio wavelengths observed in this
source. We use the observed annual cycle to constrain the velocity of
the scattering material, and the angular size 
of the scintillating component of PKS~1257$-$326.  We observe a time
delay, which also shows an annual cycle, 
between the similar variability patterns at the two frequencies. We
suggest that this is caused by a small ($\sim 10\,\mu$as) offset
between the centroids of the 4.8 and 8.6\,GHz components, and may be
due to opacity effects in the source.  The statistical properties of
the observed scintillation thus enable us to resolve source structure
on a scale of $\sim 10$ microarcseconds, resolution orders of
magnitude higher than current VLBI techniques allow. General
implications  of IDV for the physical properties of sources and the
turbulent ISM are discussed.

\end{abstract}


\section{Introduction}

Some flat spectrum extragalactic radio sources display variability on
time-scales shorter than a day. This phenomenon is known as intraday
variability (IDV), and has been studied in detail at cm-wavelengths
for two decades (\citealt{hee84,wit86}; see review by
\citealt{ww95}). There is now a great deal of evidence that
interstellar scintillation (ISS) is the principal 
mechanism causing IDV to be observed at cm wavelengths
(\citealt{jau2000,dtdb2001}, \citeyear{dtdb2002}; \citealt{ric2001},
\citeyear{ric2002}; \citealt{jm2001}). 
In turn, ISS can be
used to probe microarcsecond ($\mu$as) source structure and properties
of the Galactic interstellar medium (ISM) \citep{mj2002,ric2002}.

PKS~1257$-$326 is a flat-spectrum, radio and X-ray quasar at redshift
$z=1.256$ \citep{per98}. Its Galactic coordinates are 
$l=305.2^{\circ},~b=29.9^{\circ}$  and ecliptic coordinates
$l_{\rm ecl}=207.3^{\circ},~b_{\rm ecl}=-24.3^{\circ}$.  Radio IDV was
discovered at 4.8 and 8.6\,GHz in 2000 June with the Australia
Telescope Compact Array (ATCA) during blazar monitoring observations.
These observations revealed a compact  source showing strong flux
density changes over the course of 12 hours.  IDV was confirmed in
2000 September. Here we present the results of observations of
PKS~1257$-$326 carried out during a year-long monitoring program of a
number of the known southern IDV sources \citep{ked2001,big2002} which
commenced in 2001 February.  Results for the other IDV sources will be
presented later.  The observations and results are described in
Section 2. In Section 3 the analysis of scintillation parameters and
source characteristics is presented.  Discussion of the evidence for
scintillation being the principal mechanism for radio IDV in general,
and of the connection between scintillation and intrinsic variability,
is presented in Section 4.  In Section 5 we discuss the implications
of the three known  rapid (intra-hour) scintillators for  our
understanding of cm-wavelength IDV in extragalactic sources.  We
assume $q_0 = 0.5$ and $H_0 = 65$~km~s$^{-1}$~Mpc$^{-1}$ throughout.

\section{The Observations and Results}

The ATCA flux density monitoring was carried out simultaneously at
4.80 and 8.64\,GHz during 2001 and early 2002, with observing sessions
of 48 hours, approximately every six weeks. The target IDV sources
were interspersed with a series of compact flux density calibrators,
with observations of each source approximately every hour or less,
while they were above $15^{\circ}$ in elevation.  This technique
results in well calibrated, accurate and reliable flux density
measurements.  For PKS~1257$-$326, the time interval  between
individual measurements was decreased to gain good coverage of its
rapid variability. Additional shorter sessions concentrating on this
source alone were requested and scheduled.

PKS~1257$-$326 has exhibited IDV from the time it was discovered in
2000 June. Before this time, the only other ATCA data taken on this
source were in 1995 October. We have inspected these data, and
although they consist of only two brief (1 minute) observations,
separated by $\sim 2$ hours, the two flux density measurements differ
by $\sim 40\%$; it seems the source was also showing rapid IDV at this
time. It therefore seems likely that PKS~1257$-$326 has been showing
rapid IDV for at least seven years. Figure 1 shows both the 4.8 and
8.6\,GHz ATCA flux density measurements from 2001 March 23.  The flux
density changes remarkably rapidly, from a maximum to a minimum in
less than an hour, making PKS~1257$-$326 the third such rapid IDV
source known along with PKS~0405$-$385 \citep{ked97} and J1819$+$3845
\citep{dtdb2000}.

Figure 1 shows the remarkably smooth, quasi-periodic nature of the
variability at both frequencies, which we describe by a single
time-scale parameter, defined in Section 2.1 below.  Also noticeable
is the strong correlation between the variations at 4.8 and 8.6\,GHz
that is a characteristic of many IDV sources, in particular, the other
extreme variables \citep{ked97,dtdb2000}.

Figure 1 demonstrates the advantage of using the ATCA for such flux
density measurements; at these frequencies confusion is negligible,
the source is mostly unresolved,  and the array gives effectively
continuous, dual-frequency, high precision flux density measurements.
The data displayed in Figure 1 (and subsequent figures) are averaged
over 1 minute and over all baselines.  The thermal noise for 1-minute
integrations is $\sim 1$\,mJy, neglible compared with the observed
variations.  The variation of antenna gains with elevation and time
contributes an error which is proportional to the flux density.  This
is well-determined by the calibrator source observations. For the data
published here, typical residual proportional errors (peak-to-peak)
are up to 1\% at 4.8\,GHz, and up to 2\% at 8.6\,GHz.

There is also an error contribution due to extended source structure.
Recently we obtained high-resolution, high-sensitivity VLA images  of
the source which show it has a weak, arcsecond-scale jet (Bignall et
al., in preparation). The effect of this has not yet been subtracted
from the visibilities in the data published here, however we have
determined its contribution. The extended structure adds to the
unresolved component a small amount of flux density, which varies over the
course of a day as the $(u,v)$ coverage of the ATCA varies. The
contribution is different on different baselines, but the effect tends
to be heavily smoothed out by averaging over all
baselines. Nevertheless, at 4.8 GHz the extended structure causes  an
additional, spurious ``variation'' of $\sim 5$\,mJy peak-to-peak
at 4.8 GHz, which is in fact the dominant  error at this frequency. At
8.6 GHz, the peak-to-peak variation due to extended structure  is
$\sim 2$\,mJy.

For the purposes of estimating the time-scale and amplitude of
variability from the light-curves, the total errors are small compared
to the rapid variations, of which the {\it rms} is, on average, 16 mJy
or 8\% of the mean total flux density at 4.8 GHz, and 13 mJy or 5\% of
the mean total flux density at 8.6 GHz.  Finally, the flux density
scale is set through repeated observations of the ATCA primary
calibrator PKS~1934$-$638 \citep{rey94}. Thus, the mean flux density
of PKS~1257$-$326 between different epochs can be compared with better
than 2\% accuracy.

\subsection{Annual Cycle in the Characteristic Time-scale}

Figure 2 presents nine of the light curves at 4.8 and 8.6\,GHz made at
approximately six week intervals over the course of a year.
Immediately apparent is the dramatic change in the time-scale of the
flux density variations at both frequencies. From February through
May, the flux density varies rapidly with less than one hour between
excursions. In June the variations begin a slow down which lasts
through September. November sees them speed up again, while by 2002
January they have returned to much the same rate as 2001 February.
The time-scales estimated from the earlier discovery observations of
2000 June and September, and the more recent observations of 2002
February and April, are consistent with this {\it annual cycle} in
the  time-scale of variations.  The annual cycle provides clear
evidence that the observed rapid variations are due to interstellar
scintillation (ISS). The change in the time-scale of variations
through the year is due to the changing velocity of the scattering
medium relative to the observer  as the Earth revolves around the sun.

To quantify the observed changes we define a  characteristic
time-scale of variability.  Various authors have used different
definitions for characteristic  time-scale of IDV sources.  These are
generally based upon estimates of either the auto-correlation function
(ACF) or the structure function of the time series, or may be
estimated from the time series itself.  In studies of scintillation of
pulsars, the usual convention is to define the time-scale as the lag
at which the ACF falls to $1/e$ of its maximum value
\citep{cor86}. This definition of the ``decorrelation time-scale'',
$t_{\rm dc}$, was adopted by \citet{mj2002} in their paper on Earth
Orbit Synthesis. \citet{dtdb2000} define the time-scale by use of the
structure function  \citep[e.g.][]{sch85}.  Their definition of the
variability time-scale corresponds to the full-width  at $1/e$  of the
ACF.  \citet{ric95} and \citet{ric2002} define the time-scale as the
half-width at half maximum of the ACF, which we call $t_{0.5}$, and
which is in general slightly smaller than $t_{\rm dc}$.

For extragalactic sources, in addition to the above, several other
``counting'' parameters have been used in the literature.  This is
possible because the observed IDV is often quasi-sinusoidal,  which
leads to a deep minimum in the ACF, as shown in Figure 3.
\citet{ked97} used the mean peak-to-peak time,  while \citet{jm2001}
used the mean peak-to-trough or  trough-to-peak time. While estimates
from the ACF or structure  function are more reliable, ``counting''
methods remain useful,  particularly where the original data are
unavailable.

We calculate  the ACFs using the discrete correlation  method and
binning into time lag intervals \citep{ek88},  rather than
interpolating the data. The discrete correlation method has the
advantage that it is straight-forward to combine data from two (or
more) consecutive days without interpolating across 12 hour gaps.  The
mean is first subtracted from each data point and the ACF is
normalised to unity at zero lag. Specifically, the method is as
follows. For each pair of data points $i, j$ we calculate the lag
$(t_i - t_j)$ and the function $C_{ij}=[(S(t_i) - \bar{S})(S(t_j) -
\bar{S})]/\sigma_S^2$, where $S(t_i)$ is the flux density at time
$t_i$, $\bar{S}$ is the mean flux density for the particular dataset,
and $\sigma_S^2$ is the corresponding variance of $S$ for the
dataset. The elements $C_{ij}$ are then binned according to lag, and
the average $C_{ij}$ is taken as the estimate of the ACF for each bin,
provided a large number of elements lie within that bin.  Figure 3
shows ACFs calculated from the data of 2001 March 23.

While choosing a definition for the time scale is a matter of
convention, it is necessary to use a consistent method when  comparing
theoretical and observational results. For  PKS~1257$-$326 we have
calculated various time-scale estimates  for each frequency and for
each dataset, and find, on average, that  $t_{\rm dc} = (1.18 \pm
0.05) \times t_{0.5}$, applies  over the whole year and at both
frequencies.  For direct comparison with the results of
\citet{ric2002} and \citet{ric95}, we choose to define the time-scale
as $t_{0.5}$, the 50\% decay time of the ACF.

Figures $4a$ and $4b$ show $t_{0.5}$ determined for each session and
at both frequencies plotted against day of year, and show
quantitatively the remarkable annual cycle so apparent in Figure
2. Also included are the $t_{0.5}$ values calculated for the data from
2000, and the more recent observations of 2002 February, and April. 
The values for 2000 are plotted as limits. The light curve from June 2000
is somewhat undersampled and therefore gives an upper limit to
$t_{0.5}$, while the 2000 September data show a decrease in flux
density over 4 hours, with a subsequent 4.5 hour gap in the data (due
to observations for another project) followed by an increase over the
next 3 hours, so we have only a lower limit on $t_{0.5}$.  Plotted in
Figure $4c$  is the expected scintillation speed {\it vs} day of year,
assuming that the ISM is moving with the local standard of rest (LSR)
\citep[][and private communication]{ric2001}.  Figure $4d$ displays
the corresponding scintillation velocity projected onto the plane of
the sky.  Comparison of Figures $4a$ and $4b$ with Figure $4c$
demonstrates that the phasing of the annual cycle closely matches the
Earth's projected speed with respect to the LSR. The annual cycle in
$t_{\rm 0.5}$ is a result of the changing velocity of the
scintillation pattern across the observer's line-of-sight, and leaves
no doubt as to the origin of the intra-day variability  in this
source. Such an annual cycle has been seen in two other sources;
0917$+$624 \citep{ric2001,jm2001}, and J1819$+$3845
\citep{dtdb2001}, and is expected if the IDV is in fact a propagation
effect caused by a relatively local Galactic ``screen''. To reflect
the ISS nature of the rapid variations, we hereafter  refer to the
characteristic time-scale, estimated as the 50\% decay time of the
ACF, as $t_{\rm ISS}$.

Scintillation is a stochastic process, and since our observations
sample this process over a finite duration, it is only possible to
obtain an approximation to the true time-scale of the underlying
process.  This stochastic process, and our lack of understanding of
it, are by far the dominant contributions to the uncertainty in any
determination of the scintillation time-scale, $t_{\rm ISS}$.  The
uncertainty decreases with the number of independent samples of the
scintillation pattern, or  ``scintles'', observed.

To estimate the error empirically, we divided the datasets from 12
well-sampled epochs, when $t_{\rm ISS}$ is short and/or the dataset
extends over two consecutive days, into subsets, and calculate $t_{\rm
ISS}$ for each subset. We assume that the statistical error due to the
finite number of samples of the stochastic process scales as
$1/\sqrt{N}$, where $N$ is the number of independent samples. We
define $N$ in this case as $T_{\rm obs}/t_{\rm ISS}$, where $T_{\rm
obs}$ is the length of the observation.  We then assume that the
fractional error, $\sigma_t/t_{\rm ISS}$,  is proportional to
$\sqrt{t_{\rm ISS}/T_{\rm obs}}$. For each subset we calculate
$E_t=\sqrt{2}[t_{\rm ISS}\rm{(full~dataset)} - t_{\rm
ISS}\rm{(subset)}]$, where the $\sqrt{2}$ arises  because the two
estimates are not independent. From the observed
distribution of $E_t \sqrt{N}/t_{\rm ISS}$,  we estimate the $1\sigma$
fractional errors to be  $\sigma_t/t_{\rm ISS}=0.7\sqrt{t_{\rm
ISS}/T_{\rm obs}}$ at 4.8 GHz, and $\sigma_t/t_{\rm
ISS}=0.9\sqrt{t_{\rm ISS}/T_{\rm obs}}$ at 8.6 GHz.  The resultant
error bars are shown in Figures $4a$ and $4b$.  

To evaluate the reliability of the above error estimates, we assume
that $t_{\rm ISS}$ is constant between day of year 50 and day of year
150 (late February through May), an assumption supported by the data in
Figure 4, and by the almost constant value of $v_{\rm ISS}$ (for a
medium moving with the LSR, as shown in Figure 4$c$) over this period. 
The observed rms scatter
in the 8 data points from this period is 12\% at 4.8\,GHz and 19\% at
8.6\,GHz, compared with the mean of the errors determined for these
data as above, of 13\% at 4.8 GHz and 15\% at 8.6 GHz. This indicates that
our error estimates, as derived above, are reasonable.

Figure 2 also shows that the IDV at 4.8 and 8.6\,GHz remained strongly
correlated over 12 months of these observations; in fact we have
observed correlated variability between the two frequencies in every
observation since the discovery of IDV in 2000 June.  The increasing
difference between the mean flux density levels reveals that the
spectral index became increasingly inverted as the mean flux density
at 8.6\,GHz increased by 70\%, while that at  4.8\,GHz increased by
35\%. This is clearly seen in Figures $5a$ and $5b$, which show all
data from well-sampled epochs at both frequencies, and Figure $5c$,
which shows the mean spectral index at each epoch.  The source appears
to be undergoing a pc-scale ``outburst'' of the  type commonly seen in
many flat-spectrum AGN \citep[e.g.,][]{kpt68}. 

Figure $5d$ shows that
despite the  increase in total flux density, there is little evidence
for a systematic increase or decrease of the {\it rms} variations at
either frequency. If the increase in flux density were due to an
increase in the flux density of the scintillating component, we would
expect to see either a large increase in the {\it rms} variations, or
if the scintillating component were expanding, we may observe an
increase in the characteristic time-scale (in addition to the annual
cycle). However, neither has been observed. Regression analysis shows
that there is at best a very weak correlation in the sense that there
may be a slight increase with time of the {\it rms} variations at 8.6
GHz.  Given the low significance of this result,  we are continuing to
monitor the variations at both frequencies to look for any change in
the scintillation behaviour.  The connection between ISS and intrinsic
variability is further discussed in Section 4.

Finally, close inspection of Figure 2 reveals that at each epoch,  the
8.6\,GHz variability pattern appears to lead the 4.8\,GHz
light-curve. The time delay between the variations at the two
frequencies, $\tau_{(4.8,8.6)}$, was quantified by cross-correlating
the 4.8 and 8.6\,GHz observations at each epoch. Although the overall
variability patterns at both frequencies are quite similar, there is
some extra ``structure'' in the 8.6\,GHz light curves which appears to
be more smoothed in the 4.8\,GHz light curves. Therefore only light
curves with well-sampled peaks and troughs give a reliable time delay
measurement. As shown in Figure 6, the peak of the cross-correlation
always has the same sign, i.e. the 8.6\,GHz variations are leading.
Although the uncertainty in each individual measurement is large, due
to limited sampling and the fact that the light curves at each
frequency are not identical, this result is
significant. Interestingly, there is a clear correlation of
$\tau_{(4.8,8.6)}$ with the inverse of the predicted ISM velocity
(Figure $4c$). This gives a second annual cycle in the variability of
this  unusual source.  In the next section we  examine in detail the
implications of these results for PKS~1257$-$326 and the ISM in our
line-of-sight  to this source.

\section{Microarcsecond Source Structure}

The clear presence of an annual cycle in the time-scale of variability
identifies interstellar scintillation as the mechanism responsible for
the IDV  in PKS~1257$-$326, which in turn allows a determination of
both source and scattering screen properties.  Over the course of a
year, the Earth's orbital motion produces changes in both the
magnitude and direction of the net scintillation velocity, ${\bf
v}_{\rm ISS}$, which both potentially contribute to changes in $t_{\rm
ISS}$. Figure $4d$ shows the Earth's velocity relative to the LSR over
the course of a year, projected onto the plane transverse to the
source line-of-sight. If the scintillation pattern is isotropic, only
the change  in the nett scintillation speed contributes to variations
in $t_{\rm ISS}$.  In this case, the observed $t_{\rm ISS}$ is
inversely proportional to $|{\bf v}_{\rm ISS}|$.  However, the length
scale of the scintillation pattern may change with orientation, in
which case $t_{\rm ISS}$ will also change as the {\it direction} of
${\bf v}_{\rm ISS}$ changes. An elongated scintillation pattern could
be produced by elongated source structure and/or by anisotropic
scattering in the ISM.

\citet{ric2002} showed that the deep first minimum, which they term a
negative ``overshoot'', in the ACFs for data on PKS~0405$-$385 during
its episode of fast scintillation, indicates that the scattering is
highly anisotropic. Their results from
numerical simulation suggest that the scintillation pattern of
PKS~0405$-$385 has an axial ratio of 0.25 or less, where ${\bf v}_{\rm
ISS}$ is approximately aligned with the short axis of the
irregularities. For more highly anisotropic irregularities, the depth
of the ACF first minimum does not change significantly. The negative
overshoot is a measure of the spectral purity of the fluctuations: the
closer to sinusoidal the variations are, the closer to $-1$ is the
depth of the first minimum in the ACF.  \citet{ric2002} argue that the
negative overshoot is an effect of anisotropic scattering in the ISM,
and cannot be duplicated by anisotropic  source structure.

A large number of independent samples of the scintillation pattern are
required to determine a reliable ACF.  For light curves which sample
only 1 or 2 scintles, the ACF is poorly determined and thus a spurious
negative overshoot can occur.  For all of the long datasets on
PKS~1257$-$326, the average depths of the first minima in the
normalized ACFs are ($-0.5\pm 0.1$) and ($-0.3\pm 0.1$), at 4.8\,GHz
and 8.6\,GHz respectively. Assuming that the model of \citet{ric2002}
applies to PKS~1257$-$326, this indicates that the scattering is
anisotropic, with the {\it minor} axis of the scintillation pattern
aligned within $\sim 45^{\circ}$ of the plasma velocity vector.  The
minima at 4.8\,GHz are, on average, deeper than at 8.6\,GHz, although
there is substantial  scatter (for example, Figure 3 shows that for
the 2001 March 23 data, the first minimum of the ACF was in fact
deeper at 8.6 GHz.)  The fact that the average size of the ``negative
overshoot'' differs at each frequency  may be a result of a differing
source size relative to the Fresnel scale, i.e. the 8.6\,GHz component
may have a larger ratio of angular source size to angular size of the
first Fresnel zone, $\theta_{\rm S}:\theta_{\rm F}$, than the 4.8\,GHz
component. Hence the negative overshoot in the ACF could be suppressed
at 8.6\,GHz as source size effects become more important.

Our data indicate that the scintillations observed at 4.8 and 8.6\,GHz
are in the regime of weak scattering, where the medium introduces a
phase change of less than 1 radian across a Fresnel zone
\citep[e.g.][] {wal98}.  Correlated variability between 4.8 and
8.6\,GHz is a pattern commonly seen in other IDV sources at high
Galactic latitudes, in particular the other extreme IDVs.  At lower
frequencies,  $t_{\rm ISS}$ is generally much longer and the
variability is not correlated with the higher frequency
variability. This is explained in terms of weak and strong
scattering. At 4.8 and 8.6\,GHz, the sources undergo weak
scintillations, while lower frequencies (below $\sim 3$\,GHz) are in
the regime of strong (refractive) scattering
\citep[e.g.][]{ked97,wal98}. Time-scales of variability for refractive
scintillations increase with wavelength, $\lambda$, as
$\lambda^{11/5}$.

In the regime of weak scattering, the typical length scale of the
scintillations is equal to the Fresnel scale, $r_{\rm F} =
\sqrt{cL/(2\pi\nu)}$, where $L$ is the distance to the scattering
plasma and $\nu$ is the observing frequency. The Fresnel scale sets an
effective cut-off diameter for the angular size of a source undergoing
weak scintillation, $\theta_{\rm F} = r_{\rm F}/L$. A source this size
or smaller exhibits flux density variations on a time-scale $t_{\rm
ISS} = r_{\rm F}/{\rm v}_{\rm ISS}$, where v$_{\rm ISS}$ is the speed
of the scattering material across the line of sight.    For larger
source sizes, $\theta_{\rm S} > \theta_{\rm F}$, the scintillations
decrease in amplitude as $1/\theta_{\rm S}$ and increase in timescale
as $\theta_{\rm S}$ \citep[e.g.][]{agn:ric2002}.  It may be that AGN
have a sufficiently large angular diameter that the source size
influences the observed scintillation pattern.  Another effect of
larger source sizes is  that scintillations from distant scattering
material are suppressed, since the angular size ``cut-off'', set by
the first Fresnel zone, scales as $1/\sqrt{L}$.   Therefore, there
will be a bias towards finding sources that scintillate  behind nearby 
ISM turbulence.
%\citep{dtdb2000,ric2002}.

We can use our measurements of the annual cycle in $t_{\rm ISS}$ to
constrain (i) the velocity of the scattering medium, ${\bf v}_{\rm
ISM}$ (which Figure 4 suggests is not very different from the LSR),
(ii) the source angular size, $\theta_{\rm S}$, and (iii) the distance
to the scattering screen, $L$.   The parameters used to model the
observed annual cycle in $t_{\rm ISS}$ are as follows. ${\bf v}_{\rm
ISM}$ may be offset from the LSR and this introduces two free
parameters, the components of this offset in the plane transverse to
the source line-of-sight.  As shown in Figure 4d, the projected
velocity of the sun with respect to the LSR in RA and Dec components
is v$_{\alpha}=16.5$ km s$^{-1}$, v$_{\delta}=10.7$ km s$^{-1}$. The
RA and Dec offsets, $\delta{\rm v}_{\alpha}$ and $\delta{\rm
v}_{\delta}$, of the ${\bf v}_{\rm ISM}$ components shift the origin
in Figure $4d$. Two more parameters are required to describe the axial
ratio and orientation of anisotropy in the scintillation
pattern. Finally there is an overall scaling factor, $s_{0}$, which
sets the scintillation length scale. We choose $s_{0}$ to be the minor
axis of the scintillation pattern.  We have tested the effect of
varying these parameters, and find that the shape of the resulting
annual cycle is quite sensitive to variations in the ISM velocity and
the angle of the elongation for an anisotropic scintillation pattern.
We find the best fit parameters at each frequency by minimising the
sum of squares of the residuals.  We have examined fits both for the
case of unweighted residuals, and for the case of residuals weighted
by the error associated with each point. The shape of the slow-down
period is critical in determining the ISM velocity and anisotropy in
the scintillation pattern. However the errors in $t_{\rm ISS}$ are
larger in this period, due to having fewer independent samples of the
scintillation pattern. By using the unweighted residuals
(i.e. ignoring the errors) we allow the fit to be better constrained
by these points in the slow-down period.  On the other hand, the
overall scintillation length-scale is more accurately estimated using
the better-sampled data from the  fast period. We find that for models
which fit the data well, using weighted or  unweighted residuals makes
little difference to the results.

If we assume isotropic scattering and an isotropic source, the best
fit to the peak in $t_{\rm ISS}$ is achieved with $\delta{\rm
v}_{\alpha} = -7$ km s$^{-1}$ and $\delta{\rm v}_{\delta} = -2$ km
s$^{-1}$  at 4.8 GHz, where $\delta{\rm v}_{\alpha}$ and $\delta{\rm
v}_{\delta}$ negative have the effect of moving the origin in Figure
4$d$ further away from the ellipse.  At 8.6 GHz the best fit is for
$\delta{\rm v}_{\alpha} = -7$ km s$^{-1}$ and $\delta{\rm v}_{\delta}
= +2$ km s$^{-1}$.  Assuming the same scattering material is
responsible for the scintillation at both frequencies, we take the
average as the overall best fit: $\delta{\rm v}_{\alpha} = -7$ km
s$^{-1}$ and $\delta{\rm v}_{\delta} = 0$. These fits, and fits for
several other velocity offsets, are shown in Figures $7a$ and $7b$.
The scintillation length-scale, $s_0$, is a function of $L$ and
$\theta_{\rm S}$.   While we expect that source size influences the
scintillation in PKS~1257$-$326, at least to the extent that
scintillations from more distant scattering material are suppressed,
the size of the scintillating component, $\theta_{\rm S}$, cannot be
very much larger than the Fresnel scale at the distance of the
scattering plasma,  otherwise the source would not show such deep
modulations in flux density. The average ratio of $t_{\rm ISS}$ at the
two frequencies is $t_{\rm ISS}(4.8\,{\rm GHz})/t_{\rm ISS}(8.6\,{\rm
GHz}) =1.3\pm 0.2,$ which is consistent with being equal to
$\theta_{\rm F}(4.8)/\theta_{\rm F}(8.6)=\sqrt{8.6/4.8}=1.34.$
Therefore the source size may in fact be  smaller than, or equal to,
the Fresnel scale at both frequencies.  Alternatively, the source size
may scale with frequency in a similar way, i.e. $\theta_{\rm S}(\nu)
\propto \nu^{-0.5}$. To consider both possibilities, we introduce a
scaling parameter
$$ R_{\theta} = \cases{1, & if $\theta_{\rm S} < \theta_{\rm F}$;\cr
\theta_{\rm S}/\theta_{\rm F}, &otherwise.\cr}$$ and approximate the
scintillation length scale as $s_0 \approx L\theta_{\rm S} \approx
R_{\theta} r_{\rm F} = R_{\theta} \sqrt{cL/2\pi\nu}$. Thus
\begin{equation} \label{eq:1}
 L \approx 0.068 \left( \frac{1}{R_{\theta}^2} \right) \left(
	\frac{s_0}{10^4~{\rm km}} \right)^2 \left( \frac{\nu}{1\,\rm
	GHz} \right)~{\rm pc}
\end{equation}

Because $L \propto s_0^2$, uncertainties in $s_0$ lead to greater
uncertainties in $L$, and the screen distance is rather weakly
constrained.  For the isotropic case, the allowable range of values
found for the  scintillation length-scale are $s_0 = 6 \pm 1 \times
10^4$\,km at 4.8\,GHz, and $s_0 = 5 \pm 1 \times 10^4$\,km at
8.6\,GHz.  Letting $R_{\theta}=1$ (i.e. $s_0 = r_{\rm F}$)  gives an
approximate upper limit on $L$. Inserting these values we find $L
\approx 12 \pm 4$\,pc at 4.8 GHz, and $L \approx 15 \pm 6$\,pc at 8.6
GHz.  This implies that the scattering occurs in a very local region
of turbulence, as has been suggested for the other two intra-hour
scintillators, J1819+3845 \citep{dtdb2000} and PKS~0405$-$385
\citep{ric2002}. If $R_{\theta} > 1$, this implies an even closer
screen. For a distance in the range 10--15\,pc, $r_{\rm F}$
corresponds to an angular size in the range 30--37\,$\mu$as at
4.8\,GHz, and 22--28\,$\mu$as at 8.6\,GHz.

To obtain limits on the source brightness temperature, $T_{\rm b}$, we
also need to know the flux density, $S_c$, of the scintillating
component of PKS~1257$-$326. Following the arguments of
\citet{ric2002}, $S_c$ cannot be more than the total average flux
density and cannot be negative, so  we have $S_{\rm T} \geq S_c \geq
S_{\rm T} - S_{\rm min}$, where $S_{\rm T}$ is the total average flux
density and $S_{\rm min}$ is the lowest flux density observed. Since
we know $S_{\rm T}$ varies with time, we compute $S_{\rm T} - S_{\rm
min}$ for each epoch, and find that the maximum value is close to 50
mJy at both frequencies. Also, $S_{\rm T}$ at the start of our
monitoring was $\sim 200$\,mJy at both frequencies. In fact we can
place a further constraint on $S_c$ at 4.8\,GHz, since a
short-baseline VLBI observation of the source, including the ATCA and
an unresolved calibrator to set the flux density scale, showed that
$\sim 25$\% of the total average flux of PKS~1257$-$326 (on 2001 March
23)  is missing on an angular scale of $\sim 0.1^{\prime \prime}$ at
4.8\,GHz.  Thus we assume 50\,mJy $\leq S_c \leq$ 150\,mJy.

Assuming a circularly symmetric source brightness distribution with
FWHM corresponding to the Fresnel scale size,  an approximate lower
limit on the rest-frame brightness temperature,   $T_{\rm b} = 4
\times 10^{12}$\,K, is obtained for  $S_c=50$\,mJy and $\theta_{\rm
S}=40\mu$as at an observed frequency of 4.8\,GHz.  This requires
relativistic beaming with a Doppler factor of $\mathcal{D} \gtrsim 4$ so
as not to violate the Inverse Compton (IC)  limited brightness
temperature of $10^{12}$\,K for synchrotron   radiation.  If in the
rest frame the source is at an equipartition brightness temperature of
$\sim 2\times 10^{11}$\,K, as defined by \citet{rea94}, then
$\mathcal{D} \sim 20$ is required.  While high, these implied Doppler
factors are not  inconsistent with apparent superluminal speeds found
in VLBI surveys \citep{mar2000,kel2000}.  Taking $S_c=150$\,mJy and
$\theta_{\rm S}=30\mu$as gives a somewhat more extreme brightness
temperature of $T_{\rm b} = 2 \times 10^{13}$\,K.

Figures $7c$ and $7d$ show fits obtained allowing an anisotropic
scintillation pattern with no velocity offset from the LSR. As
discussed above, there is independent evidence for anisotropic
scattering, based on the depth of the first minimum in the
ACFs. Finally, Figures $7e$ and $7f$ show fits obtained for an
anisotropic scintillation pattern and allowing a screen velocity
offset from the LSR.  The width of the $t_{\rm ISS}$ peak, allowing
all parameters to vary,  is best fitted for both frequencies with
$\delta{\rm v}_{\alpha} = -5$ km s$^{-1}$ and $\delta{\rm v}_{\delta}
= 0$, and  when the minor axis of the scintillation pattern lies at an
angle of approximately  $25^{\circ} \pm 10^{\circ}$ to ${\bf v}_{\rm
ISM}$. The ISM velocity offset is shown by the arrow in Figure 4$d$,
and the orientation of the fitted scintillation pattern is shown by
the small ellipse in Figure 4$d$.  The best fits are found for axial
ratios $<0.5$.  Decreasing the axial ratio below $\sim 0.25$ has
little effect on the shape of the $t_{\rm ISS}$ annual cycle, because
${\rm v}_{\rm ISS}$ never cuts directly across  the long axis of the
fitted scintillation pattern.   This result is consistent with what is
expected from depth of the first minimum in the ACFs, which in the
case of PKS~0405$-$385 was fitted with a highly anisotropic medium
(axial ratio $\sim 0.25$; \citealt{ric2002}). Those sources for which
${\bf v}_{\rm ISS}$ at its maximum cuts across the short axis of the
scintillation pattern are those most likely to scintillate rapidly.

For the best fits allowing all parameters to vary, the minor axes of
the scintillation pattern scale are approximately $s_0 = 4.2 \pm 0.5
\times 10^4$\,km at 4.8\,GHz, and  $s_0 = 3.5 \pm 0.5 \times 10^4$\,km
at 8.6\,GHz.  In the case of anisotropic scattering, the typical
length scale for  the minor axis is the Fresnel scale reduced by a
factor of $\sim \sqrt{2}$ (B.J. Rickett 2002, priv. comm.).  We
estimate $L$ by setting $R_{\theta} = 1$ and replacing $s_0$  with
$r_{\rm F} \approx \sqrt{2} s_0$ in Equation~\ref{eq:1}.  This gives
values for the screen distance of $L \approx 11.5 \pm 3$\,pc at 4.8
GHz, and $L \approx 14.4 \pm 4$\,pc at 8.6 GHz; very similar values to
those obtained for the isotropic case. Hence the implied brightness
temperatures are also very similar to those found for the case of an
isotropic scintillation pattern. However, in the case of anisotropic
scattering, it may be that $T_{\rm b}$ could be reduced further if the
source was coincidentally elongated in the same direction as the long
axis of the scattering.

For rapid scintillators such as PKS~1257$-$326, it is possible to
measure the IDV pattern time delay between two widely separated
telescopes \citep{jau2000,dtdb2002}.  Recently we observed
PKS~1257$-$326 simultaneously using the ATCA and the Very Large Array
(VLA), and observed a clear time delay between the IDV pattern arrival
times at each telescope (Bignall et al., in preparation). By measuring
this delay  2 or 3 times over the course of a year, we will obtain an
accurate, independent estimate of the scintillation parameters.

\subsection{Annual cycle in the time offset between 4.8 and 8.6 GHz}

We now turn our attention to the annual cycle in $\tau_{(4.8,8.6)}$,
the offset between the IDV patterns observed at 4.8 and 8.6\,GHz. We
suggest that this may  be due to an offset between  the central
components of the source at each frequency, as might be expected if
the source were jet-like on a $\mu$as scale, and optically thick
between 5 and 8 GHz. The line in Figure 6 shows an expected annual
cycle for an  offset  between the two components, with projected
displacement vector $D(\alpha,\delta)=(1.5,0.9)\times 10^4$ km in
components of  RA and Dec, using the best-fit ISM velocity derived
from the $t_{\rm ISS}$ annual cycle above. Given the large (but
difficult to estimate) uncertainties  in the measurement of
$\tau_{(4.8,8.6)}$, this model seems a reasonable fit to the data.
The fact that $\tau_{(4.8,8.6)}$ always has the same sign means that
the ISM always crosses the 8.6 GHz ``component'' first, which severely
constrains the direction of the offset of the two components, and also
implies that the origin in Figure $4d$ cannot be inside the ellipse;
i.e. ${\bf v}_{\rm ISS}$ changes direction by  $< 180^{\circ}$ over
the year.  This is consistent with what is expected from the annual
cycle in $t_{\rm ISS}$.

For a screen distance of 10 pc, the fitted displacement vector
corresponds to an offset of $12 \mu$as.  This order of resolution is
currently unachievable by any other technique, and is comparable to
that expected for proposed future space-based X-ray \citep{cas2000}
and optical \citep{agn:unw2002} interferometry missions. An offset of
this order between different frequency components  may also have
implications for future astrometry and geodesy programs.   An angular
scale of $12 \mu$as at $z=1.256$ corresponds to a linear scale of
0.08\,pc.

Another possibility which we have not ruled out is that the frequency
offset is due to a refractive effect in the ISM rather than the source
itself. Theoretical investigation of such effects is ongoing and will
be presented in a later paper.

\section{Discussion}

The evidence for interstellar scintillation as the principal cause of
IDV at  cm wavelengths is now compelling. Three sources have shown a
clear annual  cycle in the changes in their characteristic
time-scales, J1819$+$3845,  0917$+$624, and now PKS~1257$-$326. In
addition, a time-delay has been  measured in the arrival times of the
IDV pattern at widely spaced radio  telescopes for PKS~0405$-$385
\citep{jau2000},  J1819$+$3845  \citep{dtdb2002}, and recently
PKS~1257$-$326  (Bignall et al., in preparation).

At this point, it is important to re-examine the evidence presented
for an  intrinsic origin for cm-wavelength IDV. The strongest case
reported is the correlated radio-optical changes found in 0716$+$714
\citep{qui91,wag96}. In 1990 February,  optical variations were found
that were strongly correlated with changes in the radio spectral index
between 6 and 3.6 cm wavelengths, in the sense that  the optical flux
increased as the radio spectrum became more inverted. If this
correlation were to extend across the spectrum from radio to optical
wavelengths, as would be the case if the changes were intrinsic,  then
such a spectral inversion  should lead directly to even stronger flux
density changes at  wavelengths shorter than 3.6 cm. However, such
changes are not evident in the 2 cm data for this source from the same
epoch \citep{qui2000}.  This raises some doubt as to the value of the
evidence for the intrinsic nature of the variability seen in this
source in 1990 February. It is notable that 0716$+$714 continues to
show dramatic optical variability on a time-scale of days or less
\citep{nes2002}.

Alternatively, the low level of IDV in 0716$+714$ at both 20 and 2\,cm
is consistent with a  scintillation origin, as is seen in the
variations in amplitude of IDV with  frequency in both PKS~0405$-$385
\citep{ked97} and  J1819$+$3845 \citep{dtdb2000}. Moreover, the
apparent  difference in phase between the 5 and 8\,GHz data resembles
that apparent in PKS~1257$-$326, as well as in J1819$+$3845 (see
figure 2 of \citealt{dtdb2000}), further strengthening the case for a
scintillation origin of  the radio IDV in 0716$+$714. Given its
proximity on the sky to 0917$+$624, it  is important to search for an
annual cycle in the variability time-scale of this source, similar to
that seen in 0917$+$624 \citep{ric2001,jm2001}.

The case for interstellar scintillation at cm-wavelengths  is so
strong that we suggest that this phenomenon be referred to as the {\it
physical} phenomenon of  interstellar scintillation, rather than the
{\it observational} phenomenon of IDV. This is not merely a change in
nomenclature, but has value in focusing on the physics; while IDV is
scintillation, scintillation is not necessarily IDV. For example, in
the strong (refractive) scattering regime, scintillating sources show
a characteristic time-scale which increases with decreasing frequency
($\propto \nu^{-11/5}$), that is easily an order of magnitude or more
longer than intra-day.

What is the connection, then, between scintillation and intrinsic
variability at cm wavelengths? Variability of compact, flat-spectrum
radio sources has been studied in detail for four decades. These
sources often show large ``outbursts'' on time-scales of months to
years at cm wavelengths, which have been successfully modelled as
intrinsic variations in relativistic jets, and linked with apparent
superluminal motions of components observed with VLBI
\citep[e.g.][]{kpt81}.  While refractive scintillations were found to
be important for low frequency variability (frequencies $\lesssim
1$\,GHz) \citep{hun72,ric84}, for many years the prevailing paradigm
was that cm wavelength variability was all intrinsic. It was only with
the discovery of rapid IDV \citep[e.g.][]{ked97} implying excessively
high brightness temperatures if the variability was intrinsic
($10^{21}$ K in the case  of PKS~0405$-$385), that the importance of
ISS at cm wavelengths has been extensively re-examined. In particular,
the  largest intensity fluctuations due to ISS are observed at
frequencies  close to the transition between weak and strong
scattering, typically  $\sim 3 - 8$\,GHz along lines of sight out of
the Galactic plane  \citep{wal98}.

Our flux density monitoring of PKS~1257$-$326 suggests that it is also
undergoing the type of intrinsic, pc-scale outburst commonly seen in
flat-spectrum sources, yet there has been no significant change in the
scintillating flux density. This is despite the optically thick
spectrum of the slowly varying source, and suggests that the $\mu$as
core does not lie behind what is presumably an expanding pc-scale jet,
as might be expected if the scintillating component were the core at
the base of the pc-scale jet. In other words, the outburst is probably
occurring in a separate emitting region, further out from the core in
a component too large to scintillate. High resolution VLBI
observations could provide information on activity in the source on a
scale of several pc.

It is interesting to compare the brightness temperature inferred from
the long-term intrinsic variability, with that obtained from the ISS
model. For PKS~1257$-$326, we observed a steady increase in the mean
flux density at 8.6\,GHz of $\Delta\bar{S}=0.14$\,Jy over a period of
440 days, while over the same time the mean 4.8\,GHz flux density
increased by $\Delta\bar{S}=0.08$\,Jy. The simple approach of taking
these values of $\Delta\bar{S}$ to be the flux density of the slowly
varying component at each frequency, and observed variability
time-scale  $t_{\rm int}=440$ days, gives a variability brightness
temperature (in the source proper frame),  $T_{\rm b}^{\rm var}$, of
$3\times 10^{12}$\,K  from the 8.6\,GHz long-term variability, and
$T_{\rm b}^{\rm var}=6\times 10^{12}$\,K from the 4.8\,GHz long-term
variability \citep{lv99}.  For relativistically beamed radiation with
Doppler factor $\mathcal{D}$, the inferred variability brightness
temperature $T_{\rm b}^{\rm var}$ is related to the source rest-frame
brightness temperature $T_{\rm b}^{\prime}$, by  $T_{\rm b}^{\rm
var}=\mathcal{D}^3T_{\rm b}^{\prime}$.  Where $T_{\rm b}$ is
calculated from an angular size which has been measured {\it
directly}, or estimated from an ISS model as in Section 3 above, then
$T_{\rm b}=\mathcal{D}T_{\rm b}^{\prime}$.  Taking this dependence
into account, the intrinsic  brightness temperature inferred from the
long-term variability is smaller than that calculated from the
modelled angular size of the scintillating component. This is
consistent with the long-term variability occurring in a larger
region, separate from the scintillating component, and the lack of any
detectable change in the scintillating flux density.

In some sources, variability due to ISS is episodic.  For example, the
 IDV in PKS~0405$-$385 has been convincingly shown to be due to ISS
 \citep{ked97,jau2000,ric2002}, however this source has shown only
 short-lived episodes of fast scintillation, each lasting several
 months \citep{iau182:ked2001}. PKS~0405$-$385 also exhibits slow flux
 density evolution, and \citet{ked97} proposed a possible
 interpretation of the episodic ISS, relating the long-term behaviour
 to the ejection and subsequent expansion of a component small enough
 to scintillate. However, this interpretation did not agree so well
 with the second episode of ISS observed in 1998, and it remains
 unclear whether the transient nature of the scintillation observed in
 PKS~0405$-$385 is due to intrinsic changes in the source occurring on
 time-scales of months, which cause the appearance and disappearance
 of a scintillating component, or to changes in the properties of the
 scattering medium.

A similar problem exists for 0917$+$624.  Recent monitoring data on
 this source were presented at the AGN Variability Workshop in Sydney
 by  \citet{fuh2002}, showing that the rapid, large-amplitude
 scintillation of this source has recently ceased after 15 years of
 monitoring. More recent VLA data collected during the MASIV Survey
 \citep{lov2002}, and data from the Goldstone--Apple Valley Radio
 Telescope program (M. Klein, priv. comm.) confirm that the rapid
 scintillations in 0917$+$624 have so far not returned.

While ISS and observations of annual cycles are useful for probing
$\mu$as-scale source structure,  there is clearly much remaining to be
understood about ISS of extragalactic radio sources, and its
connection to the intrinsic variability which generally occurs on
longer time-scales.

\section{Significance of the Fast Scintillators}

The three fast scintillators PKS~0405$-$385, J1819$+$3845, and
PKS~1257$-$326, are valuable because their rapid scintillation
time-scale allows one to characterize the statistics of the underlying
scintillation process over a typical 12-hour observing session; one
observes more stastically independent ``scintles'' per unit time
compared with slower scintillators. This is important because the
dominant source of error in modelling the structure of a scintillating
source is in the stochastic  properties of the scintillation process
itself.  Moreover, in their periods of rapid variability it proved
possible to make the pattern time delay measurements, crucial in
establishing the ISS origin \citep{jau2000,dtdb2002}, since this is
not possible for the more usual, slower  scintillating sources with
characteristic time scales of order a day.  These three rapidly
scintillating sources proved invaluable in making the transition  from
IDV to ISS. Viewed from an IDV perspective, their properties appear
extreme. From a scintillation perspective, however, are these sources
really so extreme?

The answer lies in the reasons for their rapid scintillation.   The
characteristic time-scale for scintillation is $t_{\rm ISS} = s_{\rm
ISS}/v_{\rm ISS}$, where $s_{\rm ISS}$ is the scintillation length
scale and $v_{\rm ISS}$ is the bulk velocity of the scattering plasma
in the observer's frame of rest.   If an annual cycle is observed in
$t_{\rm ISS}$, it follows that the scattering plasma velocity must be
close to the 30 km s$^{-1}$ of the Earth's orbital speed. As discussed
in Section 3, for an observed frequency $\nu$ in the weak scattering
regime, $s_{\rm ISS}$ is related to the Fresnel scale, $r_{\rm F}$,
which scales with screen distance $L$ as $r_{\rm F} \propto
\sqrt{L}$. $s_{\rm ISS}$ also depends on the source angular size,
$\theta_{\rm S}$, in the case where this exceeds the angular Fresnel
scale, such that $s_{\rm ISS} \sim L\theta_{\rm S}$. For most
extragalactic sources, the relation of time scale to screen distance
lies between a square root and a linear dependence \citep{ric2002}.
It follows that the fast  scintillators, those with small $t_{\rm
ISS}$, will be those scintillating sources which are to be found
behind very nearby scattering screens. The evidence in support of this
is the screen distances of $\lesssim 30$ pc found for J1819$+$3845
\citep{dtdb2000}, PKS~0405$-$385 \citep{ric2002}, and  PKS~1257$-$326.

From the scintillation perspective, these fast variables are extreme
in the  sense that they are to be found behind nearby scattering
material, not in that  there is any inherently extreme property in the
sources themselves. There is no evidence for anything intrinsically
unusual  about PKS~1257$-$326 which sets it apart from other
flat-spectrum, radio-loud quasars.  It has optical magnitude $B=18.7$
and has been detected at X-ray energies,  emitting $2\times 10^{-13}$
erg cm$^{-2}$ s$^{-1}$ between 0.1 and 2.0 keV \citep{per98}.  In
fact, its radio properties are less extreme than might be expected for
the more common, slower  scintillators. Since the angular size
required for a source to scintillate in the ISM is set by the angular
size of the first Fresnel zone, those  sources which show large
variations with longer characteristic time scales are more likely to
be found  behind scattering screens at much larger distances, and
hence with smaller  angular sizes and correspondingly higher
brightness temperatures.

There are also implications for the properties of the interstellar
medium.  The small number of fast scintillators suggests that there
are few regions  of the Galaxy where such nearby screens may be found,
a point supported by  the observation that the screen towards
J1819$+$3845 does not move at the  local standard of rest
\citep{dtdb2002}. Moreover, as  noted above, the presence of a nearby
screen relaxes the constraints on the  angular size necessary for a
source to scintillate. Thus there will likely  be more sources with
components with angular sizes of order 20 $\mu$as, which  can
scintillate through a screen at $L \lesssim 30$ pc, than there will be
sources with  5--10 $\mu$as angular sized components to scintillate
through a screen at, say, 500 pc. In this regard, it is noticeable
that the two long-lived fast  scintillators, J1819$+$3845 and
PKS~1257$-$326, are the two weakest IDV  sources reported. If the
above is correct, then a deep, unbiased scintillation  survey would be
expected to confirm the low numbers of fast scintillators.
Alternatively, the discovery of more such rapid scintillators can
reveal  the presence of such nearby scattering material in the
Galaxy. A survey is currently underway, using the VLA to look for
scintillation-induced variability over the whole Northern sky, in both
strong ($\gtrsim 0.5$\,Jy) and weak ($\sim 0.1$\,Jy) flat-spectrum
sources \citep{lov2002}.

\section{Summary}

Here we summarize the main results from our observations of
PKS~1257$-$326.
\begin{itemize}
\item
PKS~1257$-$326 is a long-lived, rapid IDV radio source with similar
properties to the other rapid IDVs, in particular PKS~0405$-$385
\citep{ked97,ric2002} and J1819+3845 \citep{dtdb2000}.
\item
Monitoring over more than a year reveals a clear annual cycle in the
characteristic timescale of variability.
\item 
This annual cycle provides unequivocal evidence for a scintillation
origin of the IDV observed in this source.
\item
Modelling the annual cycle provides strong evidence for highly
anisotropic scattering in the ISM.
\item The very rapid variability is most likely due to scintillation
in a nearby screen, only $\sim 10$ to 15\,pc from the sun.
\item
Assuming a source size equal to the Fresnel scale at the distance of
the screen, the implied brightness temperature is then $4 \times
10^{12}\,{\rm K} \lesssim T_{\rm b} \lesssim 2 \times 10^{13}$\,K,
which is high, but not entirely unacceptable for relativistically
beamed radiation from a jet.
\item 
A time offset is observed between the scintillation patterns at each
frequency, which itself exhibits an annual cycle. This we tentatively
model as an offset due to opacity effects in the source on a scale of
$\sim 10\mu$as, or $\sim 0.1$\,pc at the source.
\item
There are much slower flux density changes in the pc-scale structure
of the source, but these do not appear to be mirrored in any obvious
changes in the microarcsecond sized component responsible for the IDV.
\item
We plan to continue to monitor PKS~1257$-$326, to better constrain the
time-scale during the  slow-down period, and to follow pc-scale
changes in the source with   high-resolution VLBI observations.
\item 
For sources scintillating in the weak scattering regime which show
large variations on timescales of the order of a day, the scattering
screens may be at much larger distances (several hundred pc).  This
requires smaller source sizes, so in fact these slower scintillators
may have higher implied brightness temperatures than the very rapid
scintillators.
\item 
Observations of scintillating sources provide a method of achieving
$\mu$as resolution at cm-wavelengths with modest radio telescopes,
resolution comparable with that proposed for future space-based
optical and X-ray interferometers.
\end{itemize}

\acknowledgments

The Australia Telescope Compact Array is part of the Australia
Telescope which is funded by the Commonwealth of Australia for
operation as a National Facility managed by CSIRO.  We thank Barney
Rickett for many valuable discussions, and for code used to calculate
the expected annual change in the velocity of the scattering
screen. We also thank the referee for helpful comments. 
HEB acknowledges the support of a Faculty of Science
Scholarship from the University of Adelaide.

\begin{figure}
\figurenum{1} \epsscale{1} \plotone{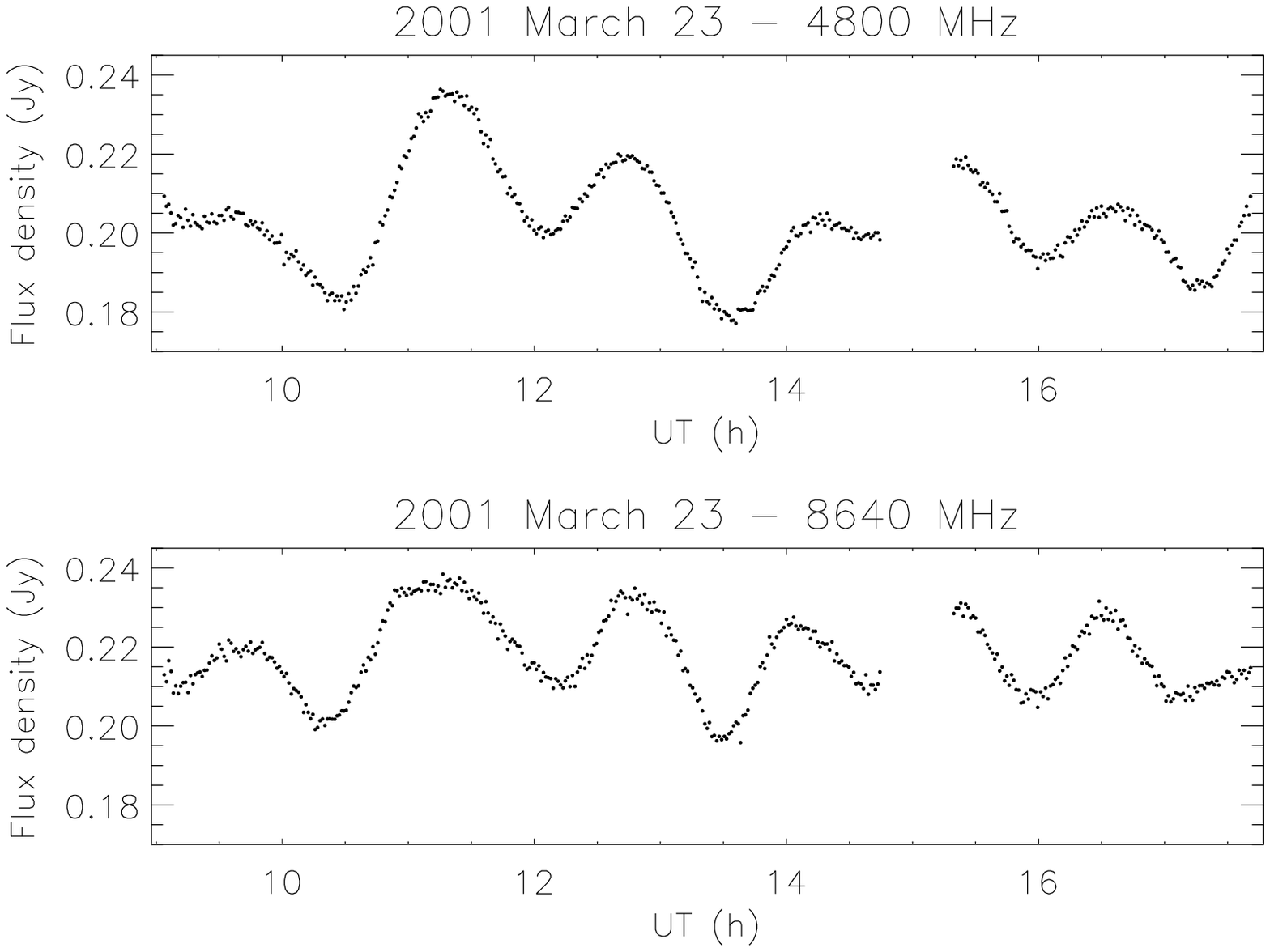}
\caption{ATCA data showing flux density variations in PKS~1257$-$326
at 4.8 and 8.6\,GHz on 2001 March 23}
\end{figure}

\begin{figure}
\figurenum{2} \epsscale{1}  \plotone{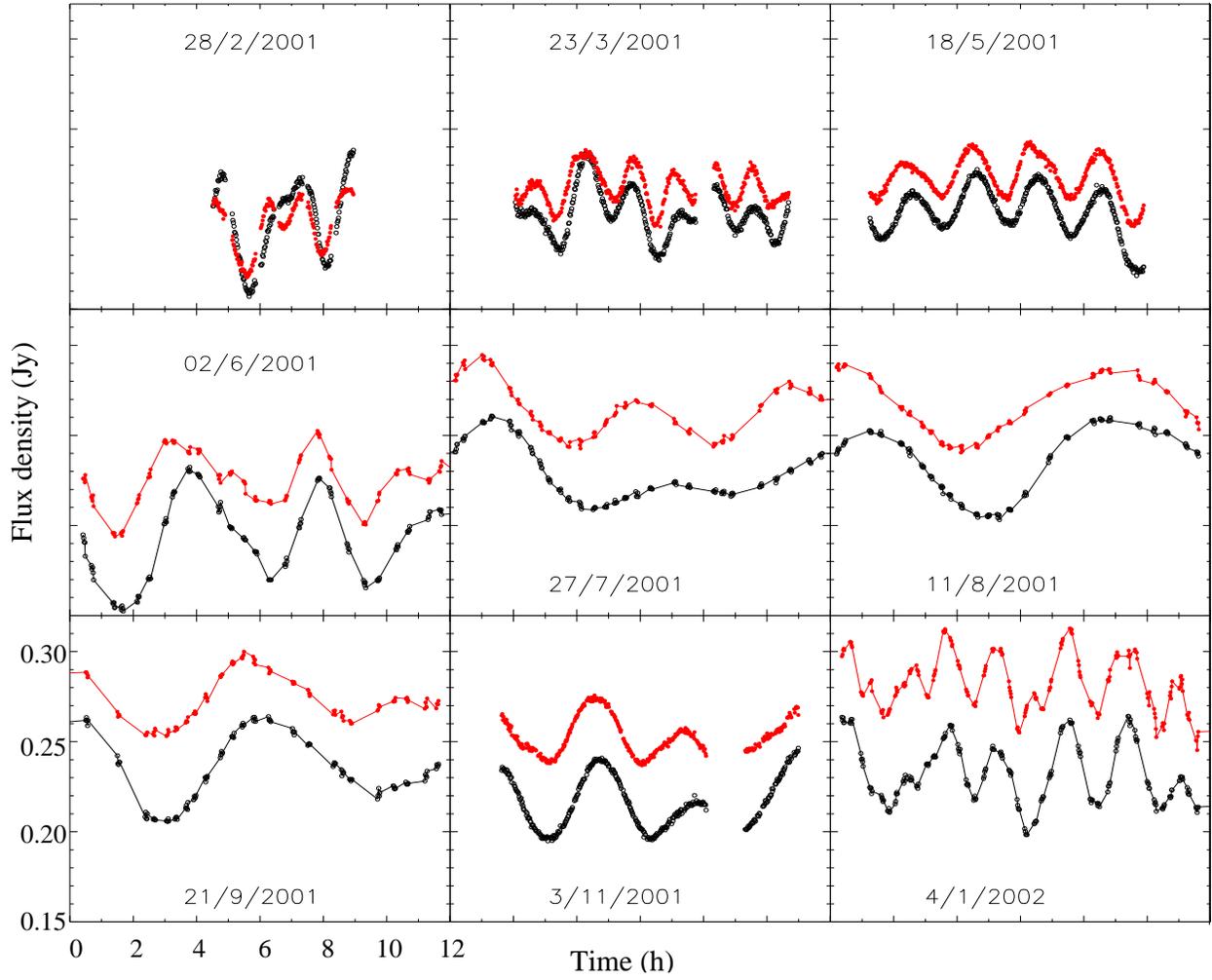}
\caption{Light curves over the course of a year for
PKS~1257$-$326. All plots are on the same scale, shown on bottom left
plot. Large symbols represent 4.8\,GHz data; small symbols (colored
red in the electronic edition) represent 8.6\,GHz data.}
\end{figure}

\begin{figure}
\figurenum{3} \epsscale{1} \plotone{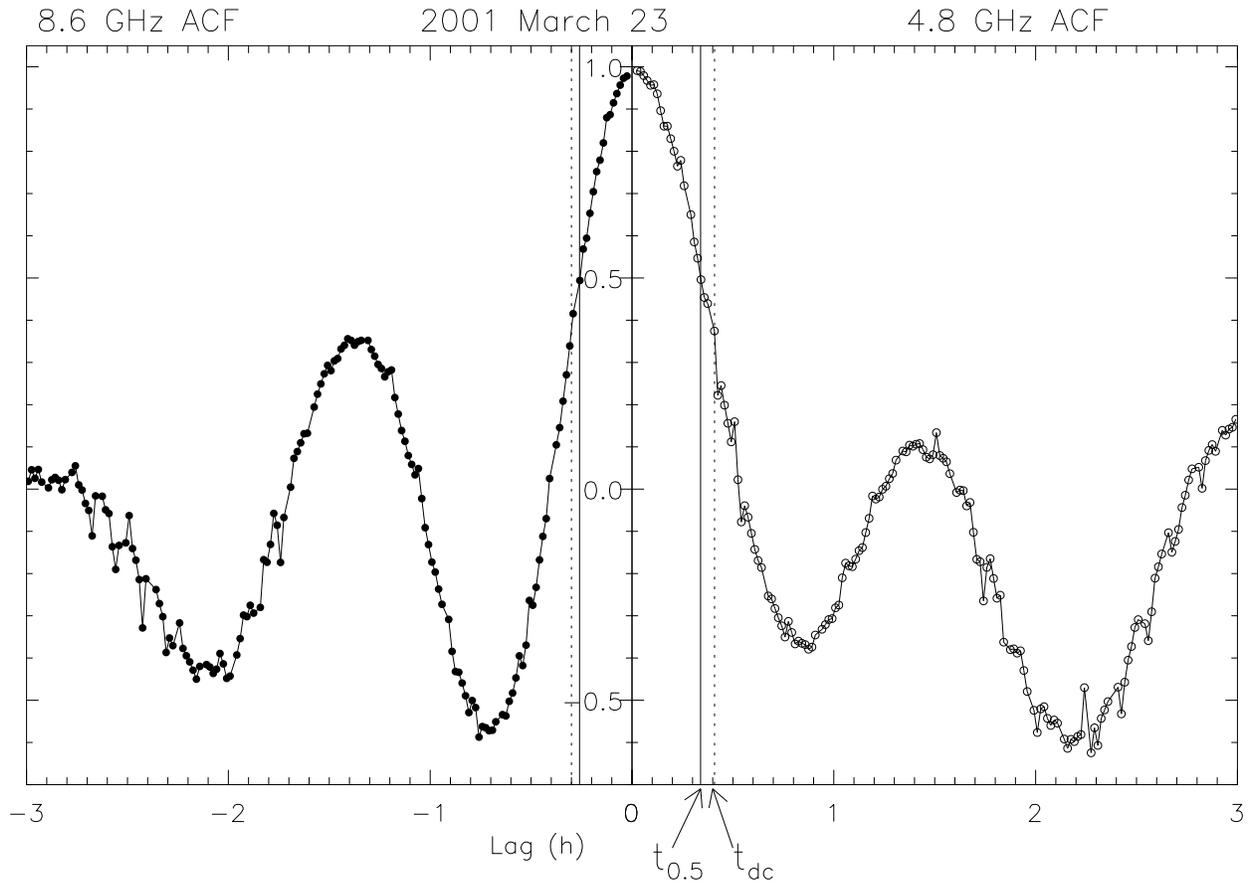}
\caption{Auto-correlation functions for data from 2001 March 23, with
common time-scale definitions shown (see text).}
\end{figure}

\begin{figure}
\figurenum{4}
\epsscale{0.8}
\plotone{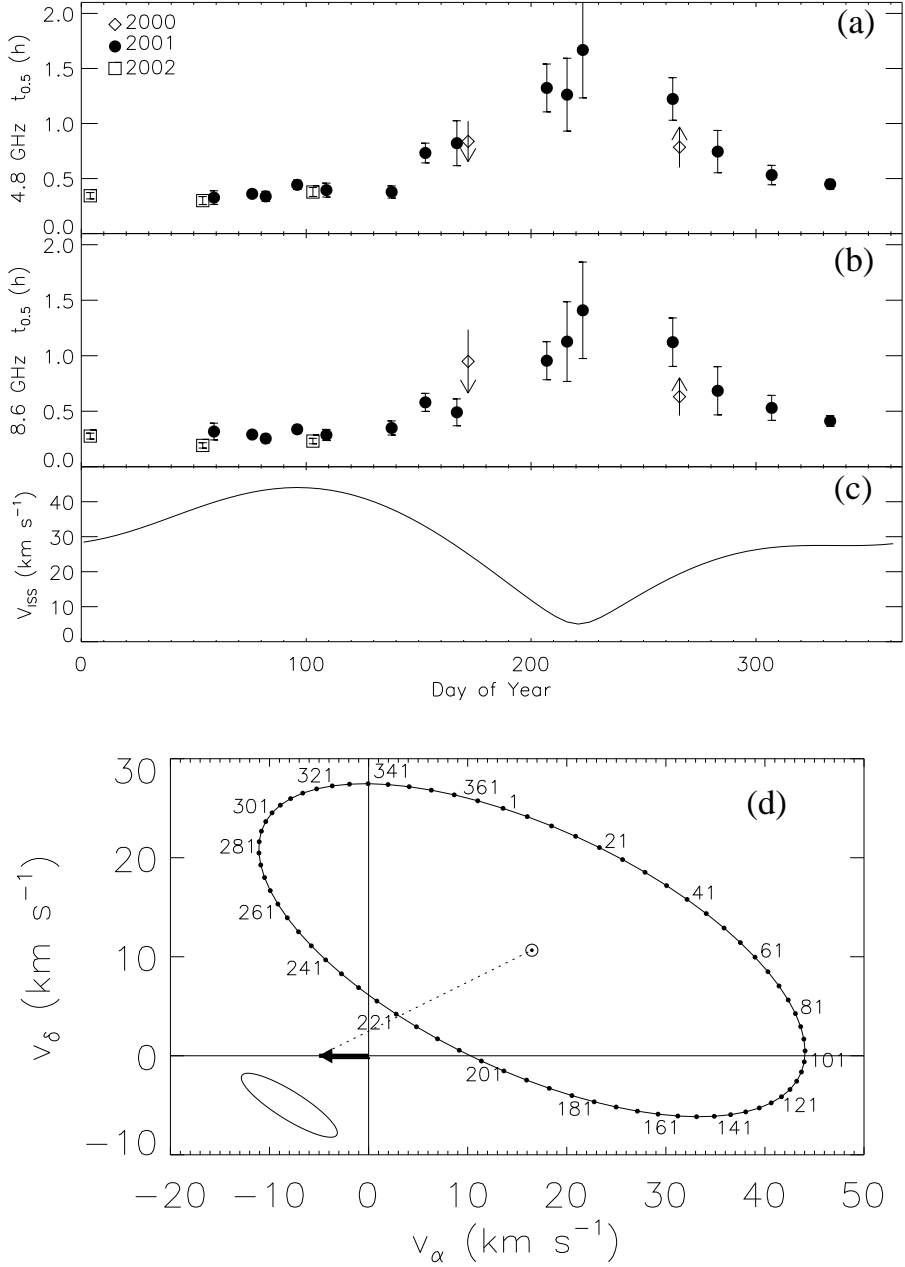}
\caption{Characteristic timescale $t_{\rm 0.5}$ (defined as the HWHM
of the ACF) {\it vs} day of year at
(a) 4.8\,GHz and (b) 8.6\,GHz. Values for 2000 are limits (see
text for details). (c) Expected scintillation speed 
v$_{\rm ISS}$ $vs$ day of year for a scattering medium moving with the LSR.
(d) The corresponding scintillation velocity ${\bf v}_{\rm ISS}$
projected onto the plane transverse to the source line-of-sight, in  
components of RA ($v_{\alpha}$) and Dec ($v_{\delta}$). 
Arrow shows offset velocity of $(\delta
v_{\alpha},\delta v_{\delta})=(-5,0)$\,km\,s$^{-1}$. Small ellipse shows
a contour of the best fit scintillation pattern. See text for details.}
\end{figure}

\begin{figure}
\figurenum{5}
\epsscale{1.0}
\plotone{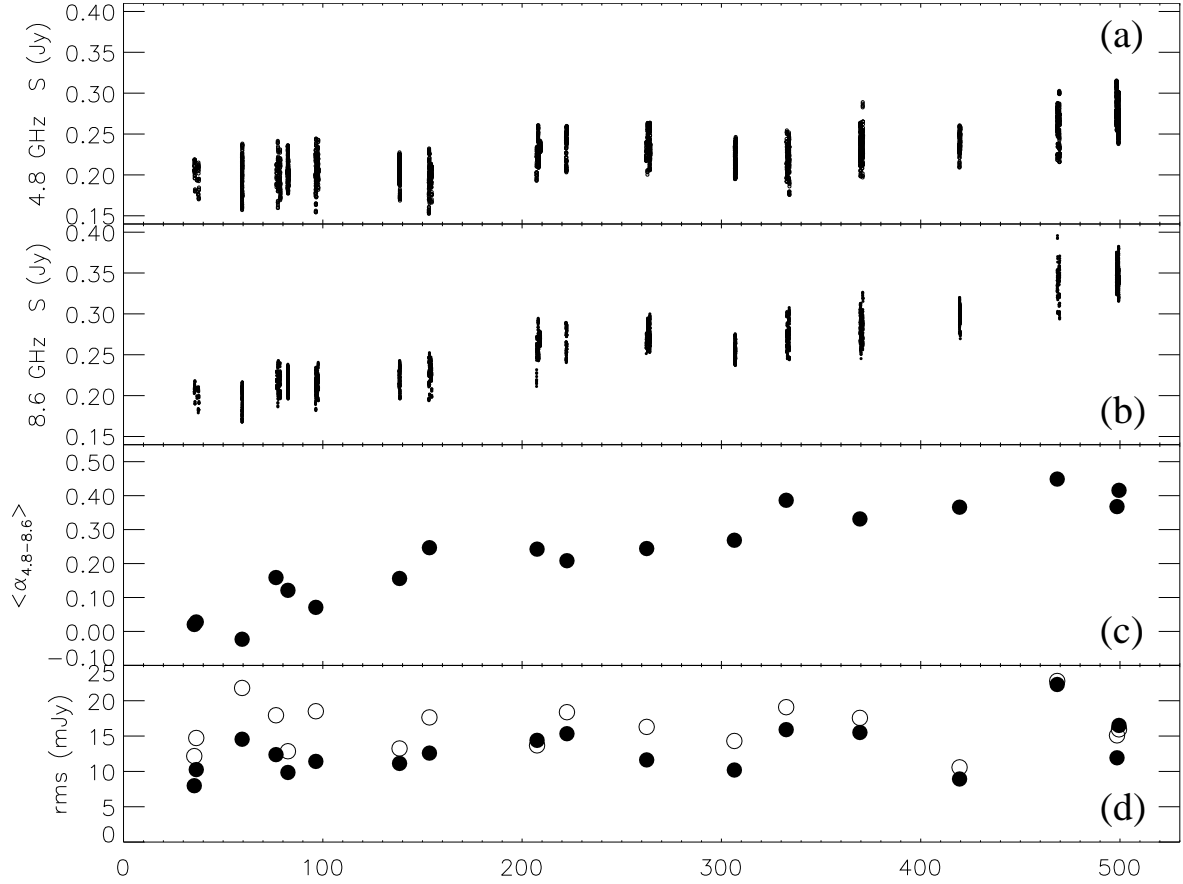}
\caption{Flux density at (a) 4.8 and (b) 8.6 GHz,  from
all well-sampled epochs, plotted with 1-minute averaging. (c) Mean
spectral index from each epoch. (d) {\it rms} variation for each epoch at
4.8 GHz (open circles) and 8.6 GHz (closed circles).}
\end{figure}

\begin{figure}
\figurenum{6}
\epsscale{1}
\plotone{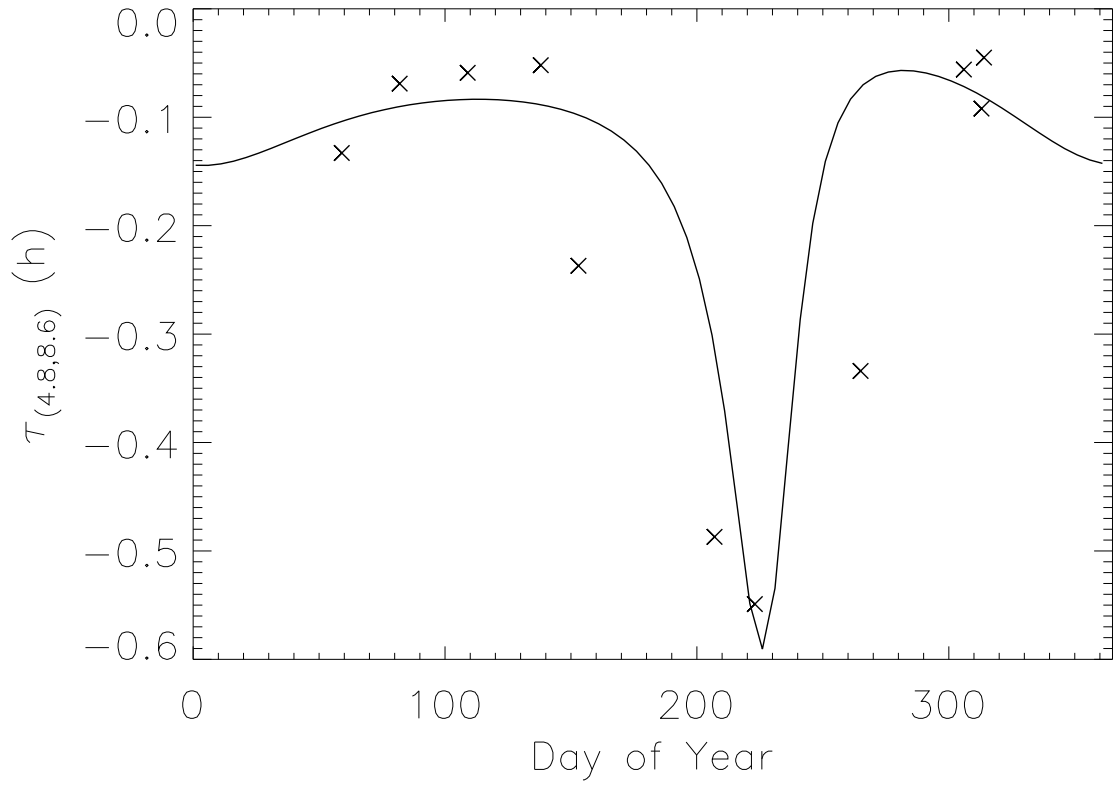}
\caption{The time offset between 8.6 and 4.8 GHz light-curves. Line
shows expected annual cycle for a fixed offset between the centroids
of the two components.}
\end{figure}

\begin{figure}
\figurenum{7}
\epsscale{0.9}
\plotone{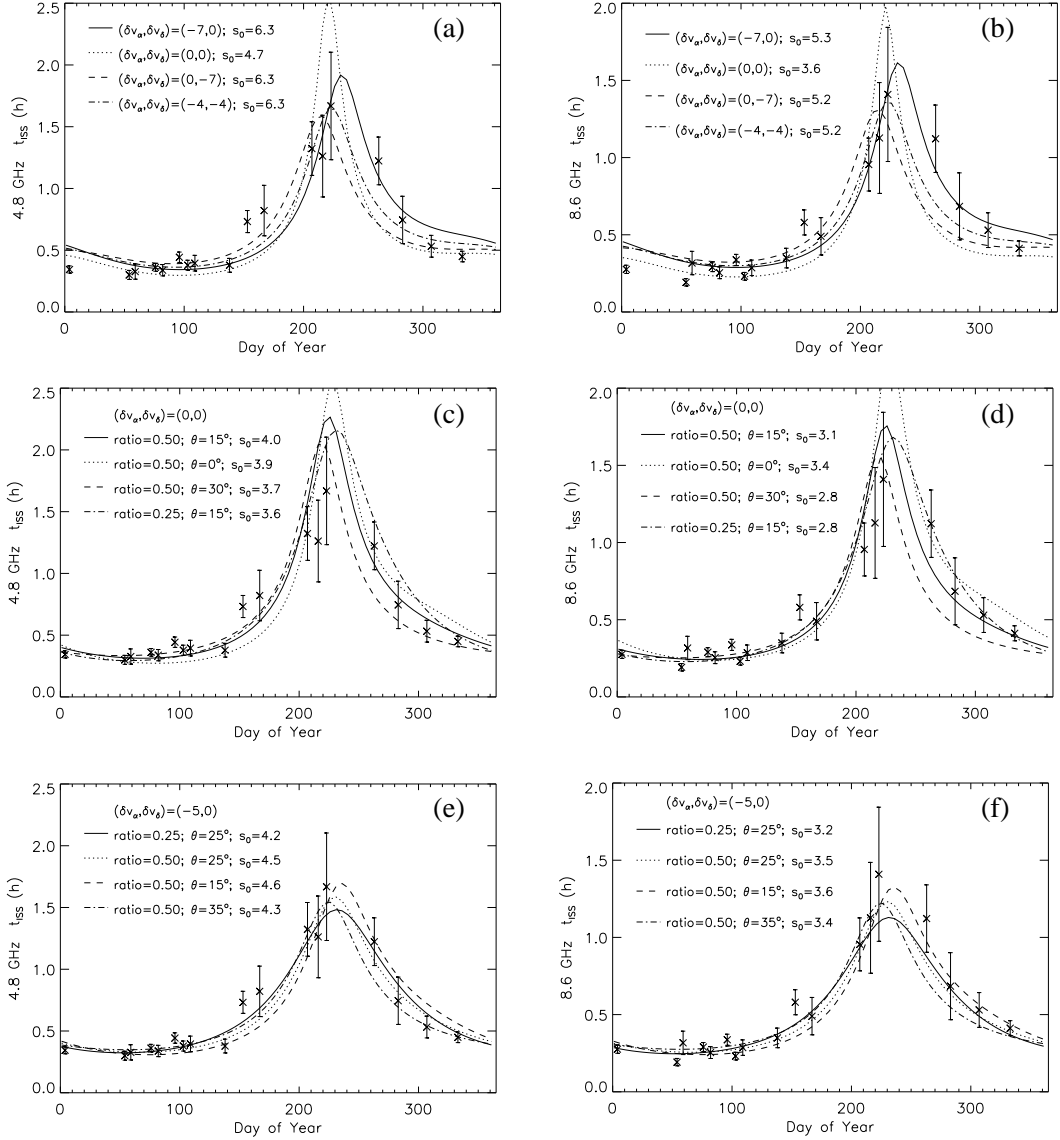}
\caption{Various fits to the annual cycle in $t_{\rm ISS}$. (a) and (b) are
for an isotropic source and isotropic scattering. (c) and (d) are for
a medium moving with the local standard of rest (LSR), allowing an
anisotropic scintillation pattern. (e) and (f) are for an ISM velocity
offset from the LSR and an anisotropic pattern. Model parameters are shown on
the plots. Ratio is the axial ratio of the anisotropy. 
$\delta {\rm v}_{\alpha}$ and $\delta {\rm v}_{\delta}$ are
in units of km s$^{-1}$, $s_0$ is in units of $10^4$ km. $\theta$ is
the angle between ${\bf v}_{\rm ISM}$ and the minor axis of the 
scintillation pattern. See text for further explanation.}
\end{figure}

\end{document}